# Mechanical identification of layer-specific properties of mouse carotid arteries using 3D-DIC and a hyperelastic anisotropic constitutive model

Pierre Badel [(1)] *, Stéphane Avril [(1)], Susan Lessner [(2)], Michael Sutton [(3)]

*(1) Center for Health Engineering,*

*Ecole Nationale Supérieure des Mines de Saint Etienne,*

*LCG – CNRS UMR5146, 158 cours Fauriel, 42023 Saint-Etienne Cedex 2, France*

*(2) Department of Cell Biology and Anatomy,*

*University of South Carolina School of Medicine, Columbia, South Carolina 29208, USA*

*(3) Department of Mechanical Engineering,*

*University of South Carolina, Columbia, South Carolina 29208, USA*

Tel: +33(0)477420260
Fax: +33(0)477499755
badel@emse.fr
http://www.cis. emse.fr/

# Mechanical identification of layer-specific properties of mouse carotid arteries using 3D-DIC and a hyperelastic anisotropic constitutive model

The role of mechanics is known to be of primary order in many arterial diseases; however determining mechanical properties of arteries remains a challenge. This paper discusses the identifiability of the passive mechanical properties of a mouse carotid artery, taking into account the orientation of collagen fibers in the medial and adventitial layers. Based on 3D digital image correlation measurements of the surface strain during an inflation/extension test, an inverse identification method is set up. It involves a 3D finite element mechanical model of the mechanical test and an optimization algorithm. A two-layer constitutive model derived from the Holzapfel model is used, with five and then seven parameters. The five parameter model is successfully identified providing layer specific fiber angles. The seven parameter model is over-parameterized, yet it is shown that additional data from a simple tension test make the identification of refined layer-specific data reliable.

Keywords: vascular mechanics ; 3D digital image correlation (3D-DIC) ; inverse identification ; hyperelasticity ; anisotropy

## 1. Introduction

Understanding the genesis and development of vascular diseases is one of the current goals of cardiovascular research. In this quest, the contribution of solid mechanics is highly relevant since many arterial disorders involve significant changes in vascular mechanical properties. The structure and mechanical response of arteries vary according to many factors such as the distance from the heart and age (Valenta et al. 1993; Fung 1973; Humphrey et al. 2003; Tremblay et al. 2010, Haskett et. al. 2010). In addition, these properties may alter under various physiological conditions and during the development of diseases. Accurate identification of mechanical and structural properties of the arteries can therefore provide helpful information for clinical diagnoses and treatments. To improve the contribution of solid mechanics, a lot of effort has been undertaken to develop constitutive models of the arterial wall as well as experimental and numerical methods to identify these models.

Several constitutive models intended to describe the mechanical response of arterial tissues at finite strains have been developed, see Vito and Dixon (2003) for an extensive review of these models. Most of anisotropic non-linear models considering the passive response of arteries are hyperelastic. Fung (1973) first introduced a phenomenological exponential strain energy function. More recently, structurally-motivated models including fiber reinforcements have been developed. Bischoff et. al. (2002) suggested an orthotropic eight-chain model to represent, in a global sense, the nonlinear orthotropic material response of the arterial wall, without distinguishing the contributions of media and adventitia. This model was further used in recent studies (Bischoff et. al. 2002, Ning et. al. 2010) considering the artery as a single homogeneous layer. It has proven adequate to capture the nonlinear orthotropic response of vascular tissues although the physical meaning of its parameters is not clear. Holzapfel et al. (2000) introduced a two-fiber family model to account for the helically-oriented distribution of collagen fibers within the arterial wall. To ideally describe the arterial wall from the mechanical point of view, two separate layers of this material are required for medial and adventitial layers. A variant of this model was proposed introducing several fiber families within one single layer (Baek et al. 2007).

All of these constitutive equations suggest various assumptions regarding the non-linear behavior of fiber families and their orientations. Originally, Holzapfel et al. (2000) grounded the choice of two symmetrically- and helically-oriented fiber families per layer on histological observations (Rhodin 1980; Canham et al. 1989; Valenta et al. 1993; Finlay et al. 1995). The reason for considering two separate layers also arose from histology of arteries since the composition of medial and adventitial layers (elastin, collagen and cell contents) are different (Humphrey 2002). The main variant of this model which is, in contrast, a one-layer model, includes additional fiber directions in axial and circumferential directions (Baek et al. 2007). It was used in several subsequent studies (Masson et al. 2008; Gleason et al. 2008; Eberth et al. 2009). Note also that some authors considered the active response of the arterial wall due to smooth muscle activity (Masson et. al. 2008), which is not the concern of this paper. These various assumptions lead to models using multiple parameters, usually ranging from five up to 14.

Correct identification of these constitutive parameters is a key issue in considering the reliability of interpretation for medical purposes or subsequent utilization in numerical models, for instance. The process of identification requires experimental data obtained from mechanical testing. When dealing with mouse carotid arteries, biomechanical testing of such small-caliber arteries is very challenging because of their small size and delicate structure. Hence, most of the previous biomechanical studies on mouse carotid arteries have been conducted using global or average data such as pressure-diameter and/or force-length measurements (Dye et al. 2007; Wagenseil et al. 2005; Guo et al. 2006; Gleason et al. 2004). The data which are used in this study are collected with a 3D-DIC stereo-microscopy system on a mouse carotid artery, which is, to our knowledge, unique at this time. See Sutton et al. (2008) for a detailed description of this previous experimental work and Ning et al. (2010) for the latest setup providing the data used here.

From experimental data, the identification of constitutive models relies, most of the time, on inverse approaches. This type of approach is necessary in many cases because establishing response curves from the model may involve complex non-linear relations between the parameters. Classical inverse approaches are based on updating methods (Holzapfel et al. 2005, Gleason et al. 2008; Masson et al. 2008, Bischoff et. al. 2009, Ning et. al. 2010) using optimization algorithms such as the Levenberg-Marquardt algorithm to find the best-fit parameters in a least-square sense with respect to a given cost function. In these approaches, previous authors have most often used analytical developments to derive their modeled data from the given constitutive equations, which presupposed multiple assumptions. Among these, note the widely-accepted assumption of axisymmetry and that of a single homogeneous layer. The latter may be relevant when experimental data is global or averaged over the arterial wall, though two separate layers would be closer to reality and would emphasize the distinction between the layer properties. In addition, for the study of mouse carotid arteries, Gleason et al. (2008) performed analytical developments within the framework of thin-tube elasticity theory. This may be a strong assumption in cases of thick arteries like mouse carotid arteries where the ratio of thickness to inner radius was reported to be about 0.6 (Dye et al. 2007; Ning et al. 2010). Note that the theory related to thick tubes was presented in (Holzapfel et al. 2000) and allows circumventing this assumption. In this previous paper, the authors identified simultaneously the material parameters of both the media and adventitia of a rabbit carotid artery using literature data, but the focus was not dedicated to the identification procedure and its reliability. To our knowledge, such simultaneous identification in both the media and adventitia has never been performed again. Identification of mechanical properties in both the media and adventitia classically entails separating mechanically the layers (Holzapfel et al.2005).

The simultaneous identification of the mechanical parameters in both the media and adventitia is one objective of the present paper.

However, in contrast to (Holzapfel et al. 2000), the present paper employs a finite element approach because it can handle more complex geometries and/or boundary effects, if any. This kind of approach has already been employed in numerous situations (Linder-Ganz et. al. 2007, Moerman et. al. 2009, Avril et. al.

2010a, Franquet et. al. 2010). To our knowledge, in the field of non linear anisotropic mechanical characterization of arteries, finite element (FE) simulations have rarely been used to recover those modeled data and perform inverse identification. Regarding non-linear anisotropic vascular properties, the study of Ning et. al. (2010) was focused on stress and strain distributions within the arterial wall and how they are influenced by axial pre-stretch. Using the same data as the present paper, they identified the parameters of the constitutive model of Bishoff et. al . (2002), thereby not considering heterogeneity between media and adventitia, which would likely affect these distributions. From our point of view, the advantage of using finite element based identification approaches is to model complex mechanical tests (Bischoff et. al. 2009, Avril et. al. 2010b) and/or complex structures, like a thick multi-layer artery presented in this paper.

The question of whether an identification method is relevant with respect to the problem to be treated is seldom addressed. Zeinali-Davarani et al. (2009) addressed the parameter estimation of multi-fiber family models for the biaxial mechanical behavior of passive arteries. Nevertheless, their interest was rather focused on the influence of measurement errors and uncertainty handling than on the modeling assumptions. Bischoff et. al. (2009) examined comparatively global and local updating techniques and confessed that the local approach, based on a Levenberg-Marquardt algorithm, did not succeed in yielding a unique solution in their problem.

Nevertheless, introducing multiple assumptions and parameters may lead to improper identification or multiple solutions. Much care has to be devoted to this kind of procedure. The objective of the present paper is to address the feasibility of the simultaneous inverse identification of mean fiber angles in both medial and adventitial layers using DIC surface strain measurements.

## 2. Methods

### 2.1. Experimental considerations

The experimental data referenced to in this paper were described in (Sutton et. al. 2008) and (Ning et. al. 2010) where three-dimensional digital image correlation (3D-DIC) is used to obtain full-field surface strain measurements on mouse carotid arteries at the micro-scale during an inflation/extension test (Fig. 1). DIC is a non-contact method based on image processing of speckle patterns before and after deformation to determine the complete full-field displacements and strains. The mechanical test performed here allows both pressurization loading and extension loading at the same time (see the schematic principle of the setup in Fig. 1). This test is interesting because of its simplicity and because it provides biaxial loading conditions. It can be noted that biaxial tests such as those developed earlier (Gleason et. al. 2004) are richer because the whole space of axial and circumferential strains can be reached by performing several tests (at fixed pressure or axial stretch), but require a more complex setup. The experiment used here presents a different way to span the space of axial and circumferential strains, with a single test, along a curve corresponding to a fixed ratio of axial to circumferential stresses.

To briefly describe this experimental setup, both ends of a freshly-dissected carotid artery are cannulated with Luer stubs. For image processing and local deformation measurements, a high contrast speckle pattern is incorporated into the vessel structure thanks to ethidium bromide nuclear staining. The experiments are performed with one end of the artery attached to the pressure controller and pressure source, while the other capped end is free in the axial direction, thereby allowing axial translation. The artery is pressurized from 5 to 150 mmHg in steps of 9 mmHg with a flow rate of 0.2 ml/min and an average pressurization rate of 1.8 mmHg/s. After each pressurization step, synchronized images are acquired from two cameras and analyzed using existing commercial software, VIC-3D.

The region of interest being small (about 200 × 140 μm²), due to the depth of field of the system, and displaying very little heterogeneity, only the strains averaged over this region are considered in the analysis. Note that usual 2D-DIC is not suitable for this problem due to the non planar nature of the specimen and possible out-of-plane deformation. More details about the experimental setup and procedure can be found in (Sutton et. al. 2008) where the method is extensively described, and in (Ning et al. 2010) regarding the data obtained and used in this paper.

*2.2. Constitutive model of the artery*

The constitutive model used in this study is based on the developments of Holzapfel et al. (2000). This hyperelastic incompressible model was developed to describe the passive mechanical response of arterial tissues at finite strains.

The material considered by Holzapfel et al. (2000) is a collagen-fiber-reinforced material with two fiber directions being symmetrically arranged with respect to the axis of the artery (Fig. 2). The theoretical basis of this model arises from composite material mechanics and relates to the mechanics of the fiber network at finite strains. This formulation provides a strong physical meaning to the constitutive parameters involved in the model. The simplest form of the isochoric strain energy function consists of two terms (note that incompressibility of the tissue is a commonly adopted assumption). The first term represents the isotropic response of the medium, related to the ground substance and elastin content, and the other two terms represent the response of the collagenous fiber network, each fiber direction having its own contribution:

$$\psi = \frac{C}{2}(I_1 - 3) + \frac{k_1}{2k_2}\left[\exp\left(k_2(I_4-1)^2\right) - 1\right] + \frac{k_1}{2k_2}\left[\exp\left(k_2(I_6-1)^2\right) - 1\right] \quad [1]$$

where:

- $C$ is the parameter of the isotropic neo-Hookean term,

- $k_1$ and $k_2$ are the parameters for the exponential response of the collagen fiber networks,

- the structural anisotropy induced by the fiber network arises from both $I_4$ and $I_6$. These terms are pseudo-invariants of the right Cauchy Green tensor $\underline{\underline{C}}$ and the fiber directions $\underline{f}_1$ and $\underline{f}_2$. Therefore they are driven by $\beta$, the mean fiber angle in the medium, defined in Fig. 2. Note that fiber angle distributions were not considered here (Gasser et. al. 2006). $I_4$ and $I_6$ give the squares of stretch for the two fiber families:

$$I_4 = \underline{f}_1 \cdot \underline{\underline{C}} \underline{f}_1 = \lambda_{\text{fiber 1}}^2 \text{ and } I_6 = \underline{f}_2 \cdot \underline{\underline{C}} \underline{f}_2 = \lambda_{\text{fiber 2}}^2 \quad [2]$$

Classically, due to histological differences between media and adventitia, separate strain energy functions are assigned to each of these mechanically-relevant layers of an artery. The contribution of the intima is commonly considered negligible. Thus, two sets of parameters are to be identified for each layer, yielding eight material parameters.

In this study, two variants of this model are considered: a simplified five parameter model and a seven parameter model. The simplified five parameter model includes the assumption that media and adventitia have identical exponential parameters. In addition, it is assumed that the value of $C$ is the same in both layers. Hence, it features the following five parameters: $C$ for the Neo-Hookean isotropic term (elastin and ground substance), two parameters $k_1$ and $k_2$ for the exponential response of collagen fibers, and two parameters $\beta^{media}$ and $\beta^{adventitia}$ for the fiber angles in media and adventitia. The corresponding strain energy function, derived from Eq. 1, is the following:

$$\psi^1 = \frac{C}{2}(I_1 - 3) + \sum_{\substack{layer = media, \\ adventitia}} \sum_{i=4,6} \frac{k_1}{2k_2}\left[\exp\left(k_2\left(I_i^{layer} - 1\right)^2\right) - 1\right] \quad [3]$$

However, a full two-layer Holzapfel-type model considers that the exponential terms used in medial and adventitial layers are different (Holzapfel et al. 2000), necessitating two additional parameters. Therefore, the seven parameter model releases the constraint on exponential parameters, making them different in each arterial layer. The corresponding strain energy function, derived from Eq. 1, is the following:

$$\psi^2 = \frac{C}{2}(I_1 - 3) + \sum_{\substack{layer = media, \\ adventitia}} \sum_{i=4,6} \frac{k_1^{layer}}{2k_2^{layer}}\left[\exp\left(k_2^{layer}\left(I_i^{layer} - 1\right)^2\right) - 1\right] \quad [4]$$

*2.3. Numerical model*

The development of the FE model of the inflation/extension test in Abaqus® is based on the experimental considerations and measurements described in Ning et al. (2010). The geometry of the artery is assumed to be perfectly cylindrical with initial inner and outer radii of 0.1003 and 0.1715 mm with a total specimen length of 2 mm. The ratio of medial thickness to total thickness is 0.45 as measured experimentally. These dimensions are similar to those measured in Dye et al. (2007). One end of the artery is capped as shown in Fig. 3. Due to axial symmetry, only one quarter of the geometry is meshed with 4280 8-node brick elements resulting in 22070 degrees of freedom. The element type chosen here, called C3D8RH in Abaqus® (hybrid formulation with constant pressure), is recommended for nearly incompressible constitutive models. The initial mesh displaying media and adventitia in different colors is shown in Fig. 3.

The open end of the cylinder is blocked in the axial direction whereas symmetry boundary conditions are applied on the surfaces of the quarter cylinder. Pressure is applied onto the inner surface of the artery, with values ranging up to 140 mmHg.

The constitutive model presented above being already built in Abaqus®, its implementation is straightforward.

The resolution of the problem is performed using an implicit scheme accounting for large displacements.

*2.4. Identification procedure*

Given a set of experimental pressure and surface strain measurements, the principle of the present identification method is to minimize the following cost function:

$$J(\vec{\chi}) = \frac{1}{2}\left[\sum_i \left(E_{11}^{sim}(p_i) - E_{11}^{exp}(p_i)\right)^2 + \left(E_{22}^{sim}(p_i) - E_{22}^{exp}(p_i)\right)^2\right] \quad [5]$$

where:

- $\vec{\chi}$ is the vector of parameters to be identified (ie. the constitutive parameters).
- $p_i$ is the pressure applied during the inflation test, with index *i* ranging over the experimental data points.
- $E_{11}$ and $E_{22}$ are Green Lagrange circumferential and axial strain components on the surface of the artery, superscripts 'sim' and 'exp' standing respectively for the simulated and experimental data. The inflation test of a long thick tube segment leads to a homogeneous strain field on the surface, provided the measurement is not performed close to the capped end of the tube. Therefore $E_{11}^{sim}$ and $E_{22}^{sim}$ are the strain components computed in a surface element far from the closed tip of the artery. $E_{11}^{exp}$ and $E_{22}^{exp}$ are the strain components obtained by averaging local strains over the region analyzed by 3D-DIC, 3D-DIC acting thus like a strain gauge. In this paper, synthetic data generated by a FE calculation are also used in order to avoid noise issues when identifying the seven parameter model.

An in-house Levenberg-Marquardt algorithm with bounds handling is used to minimize the cost function *J* defined above since this algorithm is dedicated to least-square minimization problems. Note that it requires computing the gradient of *J* with respect to $\vec{\chi}$, which is performed by backward finite differences using corresponding simulations.

The global principle of the inverse identification method is schematized in Fig. 4. Two types of stopping criteria are used: a threshold on the value of *J* (i.e. good quality of identification is reached), and another one on the norm of $\Delta\vec{\chi}$ increments (i.e. no more improvement can be reached). The first one is set to $10^{-5}$, which is very low for this problem, and the second one is set to $10^{-11}$ of $\vec{\chi}$'s norm.

To asses the robustness of the identification method, 35 identification runs with random starting points are performed in order to compare the obtained results. However in the case of the seven parameter model, since noise in data is a major source of identification errors, especially when identifying a lot of parameters, we choose to use noise-free data which are obtained by finite element simulation with an arbitrary set of parameters. Here the set of parameters obtained with the first identification run is used.

## 3. Results

In order to avoid FE convergence issues, the starting values of the minimization algorithm are chosen in the range of values found in the literature for this type of artery and for a similar constitutive model (Gleason et al. 2008).

Using the five parameter model, convergence of the optimization algorithm is obtained after 46 iterations, the stopping criterion on the norm of $\Delta\vec{\chi}$ increments being reached. We report in Table 1 the results of this identification procedure. The pressure/strain curves are shown in Fig. 5. To further test the identification method, a second set of experimental data obtained by performing a second identical test with the same arterial segment is also used, the aim being to compare the results. We also report in Table 1 the results of this identification run (curves are not shown here). Note that they are very close to those obtained with the first set of data. In the following developments, only the first set of data is used.

In addition to these results, the robustness of the identification method is assessed with the method mentioned in section 2.4. The range of spanned starting points and the range of the obtained results are reported in Table 2. Note that *C* shows quite a large standard deviation because these multiple runs showed that there exist two close minima in the space of parameters, the influence on the response being negligible.

Regarding the seven parameter model, synthetic noise-free data are generated using the first parameter set identified with the algorithm, referred to as the "true" set. Again, to evaluate the reliability of identifying this constitutive model, we try to identify its seven constitutive parameters using 35 identification random starting points. The results obtained through these tests are presented in Table 3, and the corresponding curves shown in Fig. 6. They raise several noteworthy comments.

The first comment to be made is that only three out of 35 runs lead to an agreement which is slightly not as good as most runs (cost function is found to be of the order of $10^{-5}$ versus $10^{-7}$ or less for the other runs), which means that the algorithm is practically always able to find a solution. However, only seven of 35 runs lead to the true set of parameters. The dispersion among the other results is very large; seven families of significantly different solutions can be distinguished.

## 4. Discussion

Note that residual stresses and active stress components are not considered at the moment. This paper is intended to bring a feasibility proof of the identification of layer-specific properties, and in particular mean fiber angles in each layer.

### *4.1. Relevancy of the two-layer model*

First, the authors want to emphasize the relevancy of using a two-layer model to recover the constitutive properties of this artery. This statement is not obvious, because using a single homogeneous layer model within the framework of thin-tube theory will provide a good match to the experimental data. This type of approach was used in previous studies (Gleason et. al. 2008). The analytical developments leading to identifying such a constitutive model are detailed in Appendix A. From the experimental strain measurements made at different pressures, the fiber angle thus obtained is within the range 34-35°. This range is narrow, with variations attributable to noise in the experimental data. Thus, the conclusion can be made that the constitutive parameters of the model are correctly identified with a single homogeneous layer, given the thin-tube assumption.

However, the artery considered here features dimensions which prevent from using thin-tube theory. For this reason, the procedure described in this paper, which considers the artery as a 3D thick tube modeled using the FE method, is also implemented for a single homogeneous layer. This type of model was also used in the previous work of Ning et. al. (2010) who modeled the artery's behavior with the constitutive model of Bischoff. Our results show that the best-fit parameters obtained for the constitutive model presented in section 2 are unable to capture the biaxial response of the artery (see the pressure/strain response plotted in Fig. 7). This conclusion constitutes, at least, a heuristic proof that the identification is not satisfactory when considering the actual thickness of the artery and a single layer with this constitutive model. This reasoning justifies use of the two-layer model.

Possible explanations for this result are related to 3D through-thickness effects which are not captured by the thin-tube model, while they are with the actual geometry of the artery modeled with FEM. The behavior of the thick tube is different due to triaxial effects induced by radial stresses, which disallows fitting the two exponential curves at the same time. This effect was also pointed out in Ning et. al. (2010). It can be seen from Fig. 7 that the axial strain response should be softened at low pressures only, without affecting the circumferential response. This is possible with two separate layers with different fiber angles. Indeed, the stress state varies significantly through the thickness of a homogeneous layer: though axial stress remains constant, circumferential stress shows a substantial decrease at larger radii. This classical result in elasticity theory for a thick pressurized tube was also found in Ning et. al. (2010). Therefore a lower fiber angle in the outer layer will provide a softer overall axial response while the influence on the circumferential response remains weak. The latter is mainly driven by the behavior of the inner layer where circumferential stress is highest. Our result for the five parameter two-layer model shows that balancing these effects is possible and sufficient to provide a good agreement with the axial and circumferential experimental data. This type of reasoning about stress distribution within the layers was also mentioned by Humphrey (2002) who discussed the potential interests of the inversion test in which the arterial segment is turned inside out to reverse the spatial locations of the media and adventitia.

In addition, the identification based on the second set of data and the identifications with random starting points confirm that the solution to the simplified two-layer problem is unique (see Table 1 and 2), which proves that the model is well parameterized with respect to the available data and that the method is robust.

Though residual stresses may affect the numerical values of these results and this aspect of the discussion, the question of identifiability and related problems would remain the same. However, the precise distribution of residual stresses in each layer is not known *a priori* and is difficult to hypothesize. A measure of the opening angle of the artery might help in introducing a global information on this phenomenon, but these data were unfortunately not available for the specimens used here.

*4.2. Five parameter two-layer model*

In this study, firstly, two separate thick layers are considered, with the further assumption that the material parameters in the exponential terms related to the response of fiber bundles are identical in each layer. This assumption seems to be reasonable as long as only the passive mechanical behavior is considered. The reason for this is that the passive mechanical response is mainly driven by elastin and collagen fibers. The response of elastin, as well as that of the ground substance, is included in the neo-Hookean term of the strain energy function. On the other hand, the response of collagen fibers is included in the exponential terms. Whether they are in media or adventitia, it is assumed in this model that collagen fibers of the arterial wall have the same behavior.

This parameterization of the constitutive model has been shown to be correct thanks to random starting point identifications, and to the comparison of the results obtained with the data from two different experimental tests performed on the same arterial segment.

These results confirm that identifying layer-specific fiber angles, based on the constitutive model of Holzapfel, is possible for such arteries using data from only one experimental test of inflation with free axial movement.

In our results, the mean fiber angle in the media is close to 45° while fibers are, in average, more circumferentially oriented in the adventitia. This means that the adventitia is found to be circumferentially stiffer than the media. Though it cannot be generalized based on a single example, this result brings some comments. From a global point of view the overall anisotropy of the artery, with the circumferential direction being stiffer, is typical for arteries in general, and is in qualitative agreement with previous studies, even though the constitutive models were different (Sutton et al. 2008; Gleason et al. 2008; Haskett et. al. 2010). However, the media is usually found to be circumferentially stiffer than the adventitia (Holzapfel et al. 2000, 2005), which is in contrast with the results obtained here. The possible influence of residual stresses may explain this difference and will have to be investigated in future work.

### 4.3. Additional parameters

A full two-layer Holzapfel-type model considers that the exponential terms used in each layer (media and adventitia) are different, necessitating two additional parameters (Holzapfel et al. 2000). This is the case of the seven parameter model also implemented in the present study. In contrast with the five parameter model, this model has failed in providing a unique solution to the problem of identification.

Our results (see Table 3) show the typical trend of an over-parameterized problem. They confirm that identifying such a model on these experimental data is very easy because the algorithm is always able to capture the experimental data with a close fit. Unfortunately, robustness is very poor because seven families of different possible solutions are obtained. Yet, this issue is not unexpected because the problem of inflating a tube made up of two fiber-reinforced layers potentially presents two solutions: one solution with a large fiber angle for the inner layer and small angle for the outer layer, and vice versa. Both solutions are possible as long as their constitutive properties can be different to compensate the thickness effects addressed in section 4.1. Interestingly, our results clearly show these two alternatives, with a noticeable preference for large media angle and low adventitia angle (28 out of 35 solutions). Note, also, that the mechanism of compensation is clearly illustrated by the ratio of $k_1$ parameters between media and adventitia, which is inversed in these two situations.

To correctly identify the parameters for this kind of model, more abundant and/or relevant experimental data are required. In order to confirm this hypothesis, additional synthetic data are generated by simulating, from the same FE model, an axial tension test on the arterial segment, without any pressurization, and recording the axial force versus axial stretch. The material parameters used for the generation of these synthetic data are arbitrarily chosen to be those of the second family of previous solutions (see Table3). Accordingly the cost function (see eq. 5) is enriched with these data:

$$J(\vec{\chi}) = \frac{1}{2} \left[ \sum_i \left( E_{11}^{sim}(p_i) - E_{11}^{exp}(p_i) \right)^2 + \left( E_{22}^{sim}(p_i) - E_{22}^{exp}(p_i) \right)^2 + \sum_i \left( F^{sim}(\lambda_i) - F^{exp}(\lambda_i) \right)^2 \right]$$
[6]

where:

- $\lambda_i$ is the axial stretch applied during the tension test, with index *i* ranging over the available experimental (synthetic) data points.

- $F^{sim}$ and $F^{exp}$ are simulated and experimental (synthetic) axial force. Values are scaled so that both sum terms in J are of the same order of magnitude.

Using this updated cost function, the same 35 identification runs are performed again. Instead of obtaining seven different families of solutions, as with the former cost function, the algorithm provides only two (see Table 4). Among these two solutions, the first one, obtained five times out of 35, reaches two boundaries of the allowed range for the parameters. This solution must obviously be rejected, as the procedure would have diverged or found inconsistent values. The other one, obtained 30 times out of 35, can be considered as the true solution.

These results show that additional relevant data may easily make the identification procedure robust. In this case, a simple tension test added to the inflation/extension test is sufficient. However, any larger set of data obtained with an experimental setup allowing spanning widely the strain or stress space would be helpful as well. The interest of the two tests used in the present paper is that they do not require specific biaxial equipment. If available, such equipment will obviously be of a great support.

Circumventing the problem of non-uniqueness emphasized above would also be made possible by using different types of additional data, or other strategies. For instance, separate mechanical tests of medial and adventitial layers can be used as was done in Holzapfel et al. (2005), or inversion tests which were theoretically studied in Humphrey (2002). This would be, though, very challenging with such small vessels as mouse carotid arteries. Another possible way to access useful experimental information would be to acquire through-thickness data, thanks to the use of optical coherence tomography or confocal microscopy, for instance. Such layer-specific data would help discriminate the true solution. Otherwise, in order to reduce the number of dependent parameters to be identified and help make the solution unique, additional information regarding the fiber angle distributions within each layer (from histology for instance) would be helpful. A part of the work in progress at this time is related to these last two aspects.

### *4.4. Symmetry of the structure*

The assumption of geometric and constitutive axisymmetry of arteries is widely accepted and used in the literature. The inflation/extension tests performed on those mouse carotid arteries raise, though, the question of axisymmetry. Indeed, full-field strain measurements have reported significant shear strains occurring during the test. Values up to 3 % have been recorded, which may make the assumption of axisymmetry questionable.

One can imagine two ways of including such effects in the model in order to identify additional parameters related to this behavior. The first one would be to consider that the twist of the artery is a consequence of an imbalance in fiber directions. Then considering that the fiber directions are not symmetrically oriented would lead to identification of two fiber angles per layer instead of one. The second approach would be to consider that shear results from an imbalance in fiber proportions in each direction. This would lead to introduction of a multiplying factor between the responses in each fiber direction.

These perspectives require further developments of the model, therefore they are part of future work.

### 5. Conclusion

In this study, 3D-DIC is used to measure surface strains on the surface of a mouse carotid artery. Instead of the usual pressure-diameter measurements for which post-analysis often suffers approximations such as the thin-tube assumption, the actual surface strains are acquired. The objective of the present paper was to study the feasibility of recovering fiber angles in each arterial layer from these data.

A Holzapfel-type material model is identified using an inverse method based on a finite element model and a Levenberg-Marquardt optimization algorithm. Layer-specific mean fiber angles are thus determined using a five parameter constitutive model demonstrating good robustness of the identification procedure. Importantly, we show that a model based on a single thick layer is unable to render the biaxial mechanical response of the artery tested here. On the contrary, difficulties related to the identification of a seven parameter constitutive model are evidenced; such a model leads to multiple solutions. Nevertheless, it is shown that an additional mechanical test, different in nature with the previous one, solves this problem. Additional data such as through-thickness or histological data are also likely to be the key to such refined models. Current work is dedicated to improving this aspect.

**Appendix A**
In this appendix, the analytical developments made to recover the constitutive properties of a one-layer thin-tube model are detailed.

Let $\{\underline{e}_i\}$ denote the reference frame as a circular permutation of the cylindrical frame linked to the artery: $\underline{e}_1 = \underline{e}_\theta$ ; $\underline{e}_2 = \underline{e}_{zz}$ ; $\underline{e}_3 = \underline{e}_{rr}$. Assume the principle axes of the transformation are aligned with the axes of the reference frame $\{\underline{e}_i\}$, and assume no shear. Note that this is the case of the inflation/extension test considered here. The gradient tensor is thus diagonal and the following kinematical tensors can be written:

$$[\underline{\underline{\mathbf{F}}}]_{\{\underline{e}_i\}} = \begin{bmatrix} \lambda_1 & 0 & 0 \\ 0 & \lambda_2 & 0 \\ 0 & 0 & \lambda_3 \end{bmatrix}$$

$$[\underline{\underline{\mathbf{C}}}]_{\{\underline{e}_i\}} = \begin{bmatrix} \lambda_1^2 & 0 & 0 \\ 0 & \lambda_2^2 & 0 \\ 0 & 0 & \lambda_3^2 \end{bmatrix}$$

$$[\underline{\underline{\mathbf{E}}}]_{\{\underline{e}_i\}} = \frac{1}{2}\begin{bmatrix} \lambda_1^2-1 & 0 & 0 \\ 0 & \lambda_2^2-1 & 0 \\ 0 & 0 & \lambda_3^2-1 \end{bmatrix}$$

According to Fig. 2, let $\underline{\mathbf{f}}_1$ and $\underline{\mathbf{f}}_2$ denote the fiber directions oriented at angles $\pm \beta$ with respect to the direction $\underline{e}_1$. The stretch of the fibers can be expressed as a function of principle stretches or, equivalently, as a function of Green Lagrange strains:

$$I_4 = I_6 = \lambda_1^2 \cos^2 \beta + \lambda_2^2 \sin^2 \beta = 2E_{11}\cos^2\beta + 2E_{22}\sin^2\beta + 1$$

Using Holzapfel's model with a single homogeneous layer for the whole artery wall (and neglecting the neo-Hookean term for the sake of simplicity), the following second Piola-Kirchhoff stress components can be derived from Eq. 1:

$$S_1 = \frac{\partial \psi}{\partial E_1} = 4k_1 \cos^2\beta (I_4-1)\exp\left(k_2(I_4-1)^2\right)$$

$$S_2 = \frac{\partial \psi}{\partial E_2} = 4k_1 \sin^2\beta (I_4-1)\exp\left(k_2(I_4-1)^2\right)$$

Cauchy stresses are then given by:

$$\underline{\underline{\sigma}} = \frac{1}{J}\underline{\underline{\mathbf{F}}}.\underline{\underline{\mathbf{S}}}.\underline{\underline{\mathbf{F}}}^T \Rightarrow \begin{cases} \sigma_1 = 4k_1\lambda_1^2 \cos^2\beta(I_4-1)\exp\left(k_2(I_4-1)^2\right) \\ \sigma_2 = 4k_1\lambda_2^2 \sin^2\beta(I_4-1)\exp\left(k_2(I_4-1)^2\right) \end{cases} \quad [\text{A.1}]$$

Within the assumption of thin-tube theory, the following relationships can easily be retrieved for the inflation/extension test:

$$\begin{aligned} \sigma_1 &= \frac{pR}{t} \\ \sigma_2 &= \frac{pR}{2t} \end{aligned} \quad [\text{A.2}]$$

where $p$ is the inner pressure, $R$ the mean radius of the thin tube and $t$ its thickness. Then, using Eq. A.1 and A.2 it should be verified, for each pressure level, that:

$$\frac{p\,R}{t} = 4k_1\left(I_4 - 1\right)\exp\left(k_2\left(I_4 - 1\right)^2\right)\lambda_1^2 \cos^2\beta$$

$$\frac{p\,R}{2t} = 4k_1\left(I_4 - 1\right)\exp\left(k_2\left(I_4 - 1\right)^2\right)\lambda_2^2 \sin^2\beta$$

This expression immediately yields the fiber angle $\beta$:

$$\tan\beta = \frac{\lambda_1}{\lambda_2\sqrt{2}}$$

Parameters $k_1$ and $k_2$ are subsequently obtained by fitting the exponential pressure/strain response.

**List of figures**

Figure 1. Schematic of the inflation/extension test of the mouse carotid artery showing one fixed end (on the left) linked to the pressure controller and one end (on the right) free to translate axially. Dashed lines represent the schematic shape of the deformed segment.

Figure 2. Schematic of an arterial layer as considered in the Holzapfel constitutive model ; fiber angle β is defined with respect to the circumferential direction $\underline{e}_1$ and axial direction $\underline{e}_2$

Figure 3. Initial geometry and mesh of the artery used in the FE model of the inflation/extension test. Different colors are used for the medial and adventitial layers.

Figure 4. Flow chart of the inverse identification procedure

Figure 5. Pressure/strain curves obtained with the five parameter constitutive model.

Figure 6. Pressure/strain curves obtained with the seven parameter constitutive model.

Figure 7. Pressure/strain curves for the one-layer Holzapfel model (4 parameters).

**List of tables**

Table 1. Material parameters determined by inverse identification of the simplified two-layer Holzapfel model. Two different experimental data set obtained with the same arterial segment have been used for comparison.

|  | C (kPa) | k1 (kPa) | k2 | $\beta^{media}$ (°) | $\beta^{adventitia}$ (°) |
|---|---|---|---|---|---|
| Starting point | 20 | 20 | 1 | 35 | 35 |
| First data set | 0.5 | 33 | 12.8 | 46.4 | 27.2 |
| Second data set | 0.86 | 31.3 | 14 | 43.3 | 27.3 |

Table 2. Results of parameter identification of the five parameter model with random starting points. Results are presented as mean values ± standard deviation.

|  |  |  |  | Media | Adventitia |
|---|---|---|---|---|---|
|  | $C$ (kPa) | $k_1$ (kPa) | $k_2$ | $\beta$ (°) | $\beta$ (°) |
| Starting point range | [0.3;50] | [0.5;100] | [0.1;100] | [5;60] | [5;60] |
| Results range | 0.45±0.37 | 30.5±1.07 | 15.4±0.51 | 46.7±0.2 | 26.8±0.17 |

Table 3. Results of the identification of the seven parameter model with random starting points. The different families of solutions obtained, and their occurence are presented.

|  |  |  | Media |  |  | Adventitia |  |  |
|---|---|---|---|---|---|---|---|---|
|  |  | $C$ (kPa) | $k_1$ (kPa) | $k_2$ | $\beta$ (°) | $k_1$ (kPa) | $k_2$ | $\beta$ (°) |
| Starting point range |  | [0.3;50] | [0.5;100] | [0.1;100] | [5;60] | [0.5;100] | [0.1;100] | [5;60] |
| True parameters |  | 8.8 | 13.9 | 21.6 | 41.9 | 11.8 | 0.242 | 5.14 |
| Solution family | Occurrence (%) |  |  |  |  |  |  |  |
| n°1 | 40 | 1.01 | 0.01 | 52 | 55.1 | 15.1 | 17.4 | 7 |

| | | | | | | | |
|---|---|---|---|---|---|---|---|
| n°2 | 20 | 8.8 | 13.9 | 21.6 | 41.9 | 11.8 | 0.31 | 5.4 |
| n°3 | 14.3 | 0.75 | 0.8 | 0.1 | 5 | 19.6 | 20.5 | 38.8 |
| n°4 | 11.4 | 8.46 | 13.6 | 21.7 | 41.8 | 14.5 | 0.1 | 15.5 |
| n°5 | 8.6 | 18.1 | 22.9 | 27 | 53.7 | 43.8 | 14.4 | 30 |
| n°6 | 2.8 | 13.4 | 14.9 | 13.7 | 35.2 | 87 | 58 | 12.9 |
| n°7 | 2.8 | 8.3 | 10.2 | 5 | 7.4 | 8.8 | 100 | 51.9 |

Table 4. Results of the identification of the seven parameter model with random starting points, using the cost function enriched with tensile test data. The different families of solutions obtained, and their occurence are presented.

| | | Media | | | | Adventitia | | |
|---|---|---|---|---|---|---|---|---|
| | | $C$ (kPa) | $k_1$ (kPa) | $k_2$ | $\beta$ (°) | $k_1$ (kPa) | $k_2$ | $\beta$ (°) |
| Starting point range | | [0.3;50] | [0.5;100] | [0.1;100] | [5;60] | [0.5;100] | [0.1;100] | [5;60] |
| True parameters | | 8.8 | 13.9 | 21.6 | 41.9 | 11.8 | 0.242 | 5.14 |
| Solution family | Occurrence (%) | | | | | | | |
| n°1 | 85.7 | 8.8 | 13.7 | 21.8 | 42 | 12.3 | 5.4 | 8 |
| n°2 | 14.3 | 8.8 | 8.11 | 0.1 | 5 | 14.2 | 64.7 | 38.9 |

Figure 1:

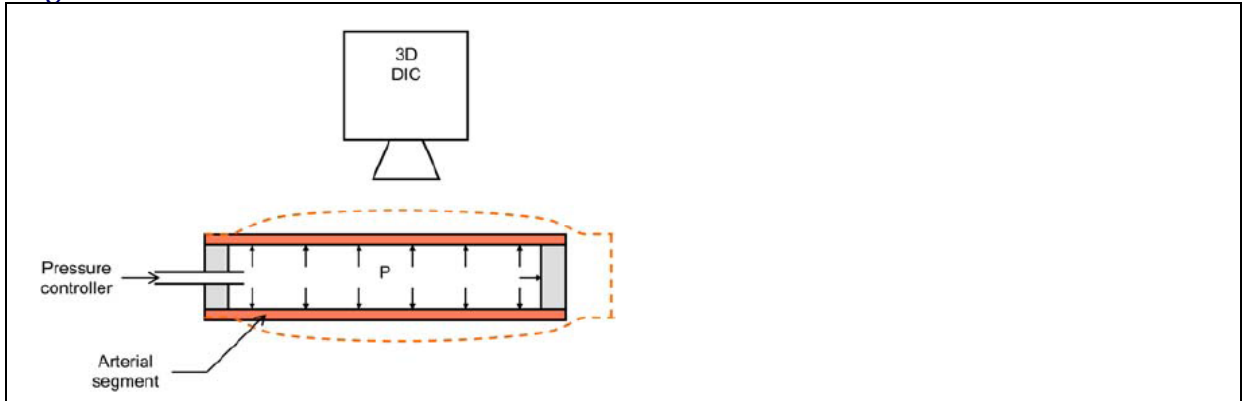

Figure 2:

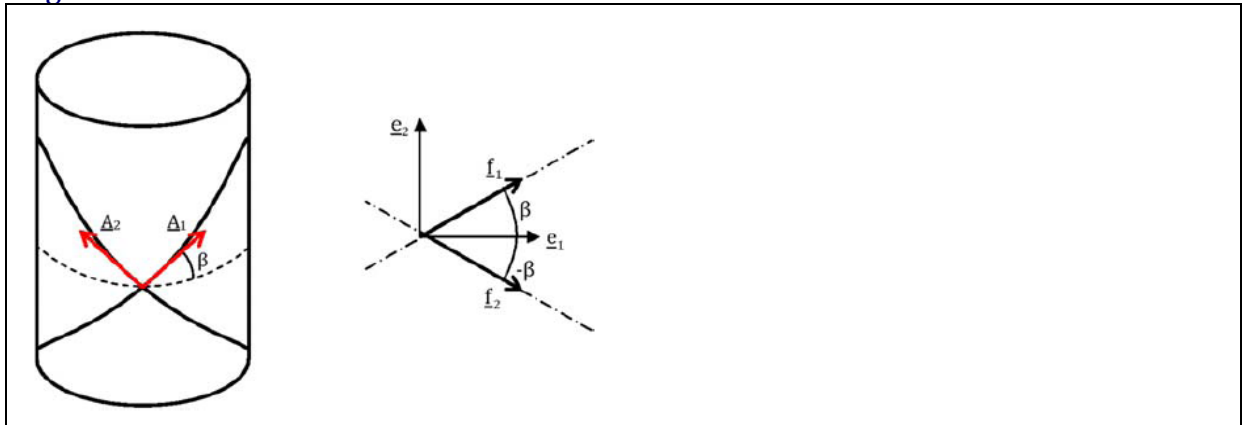

Figure 3:

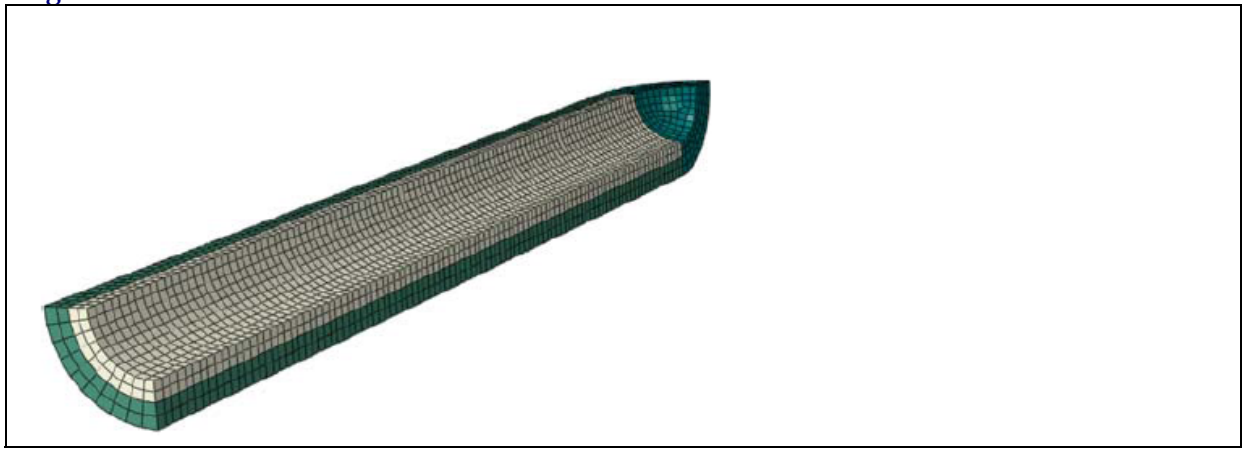

Figure 4:

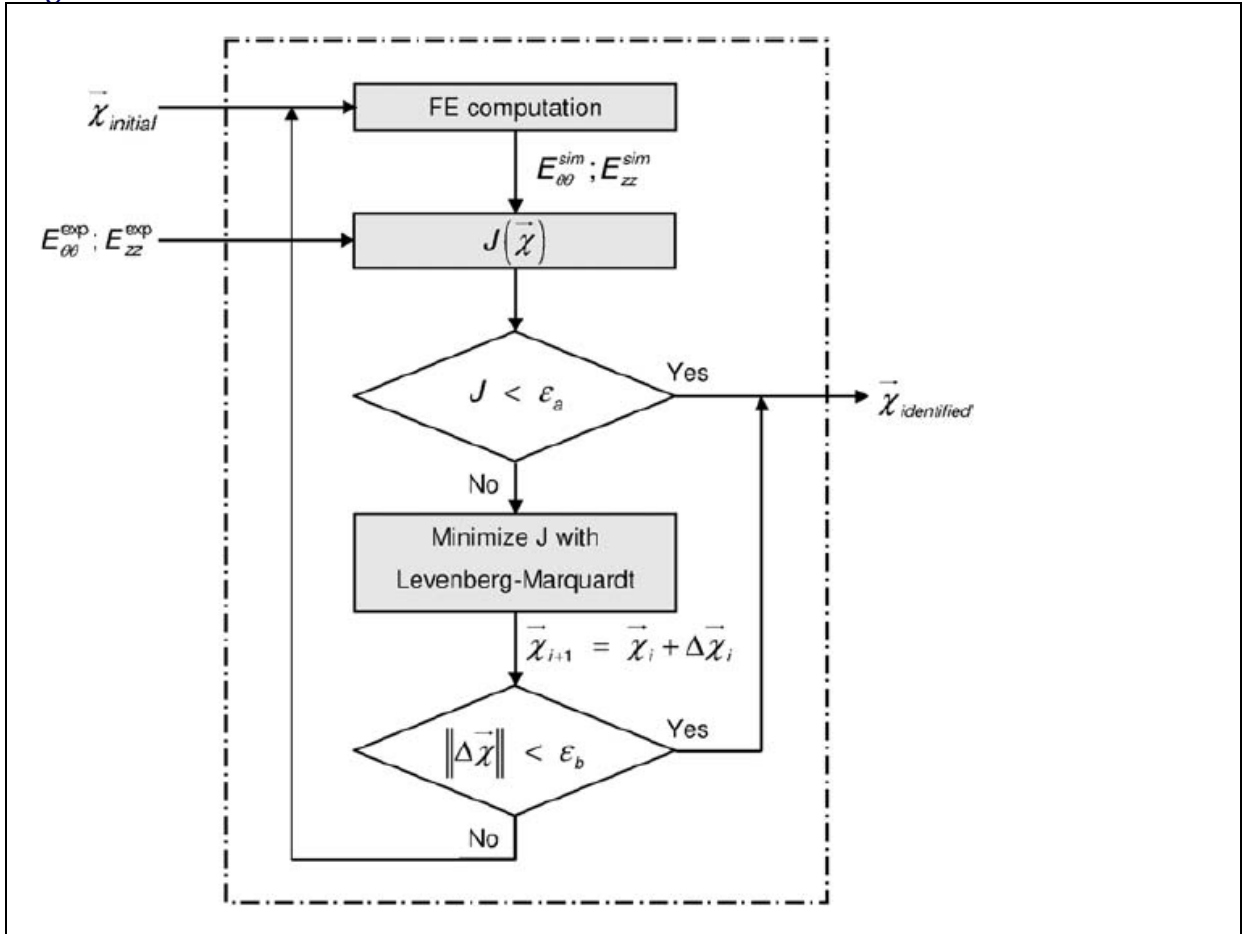

Figure 5:

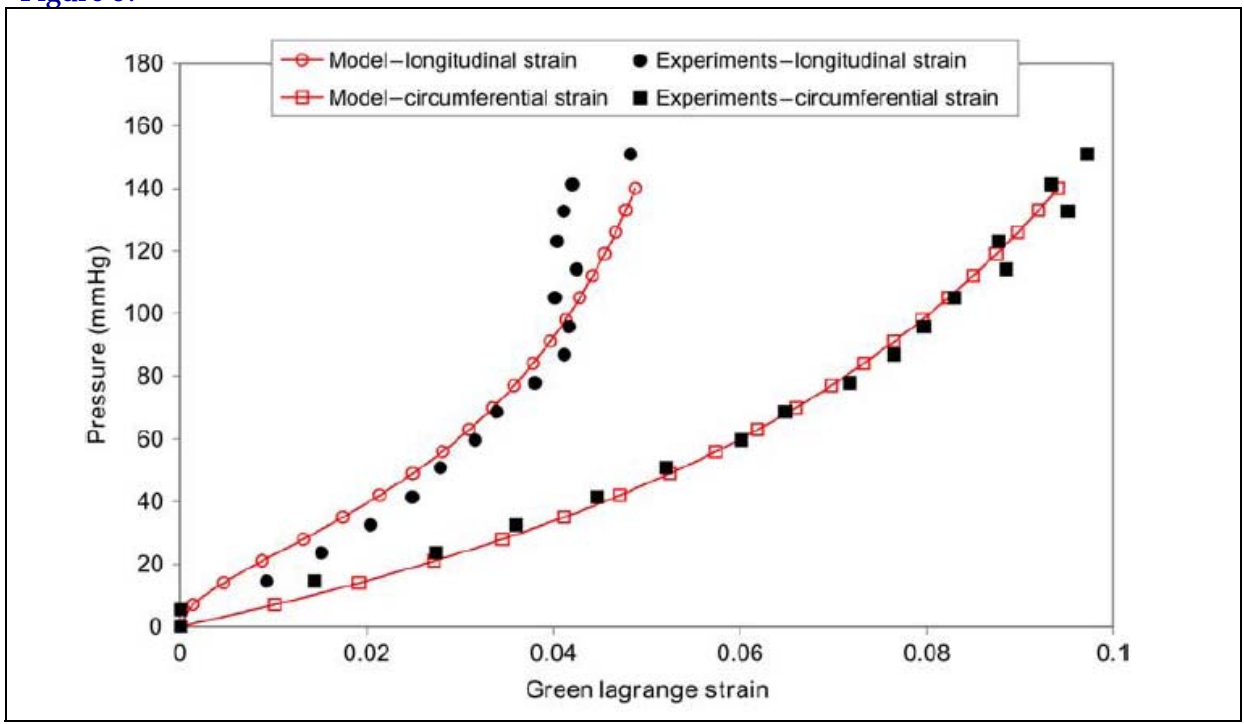

Figure 6:

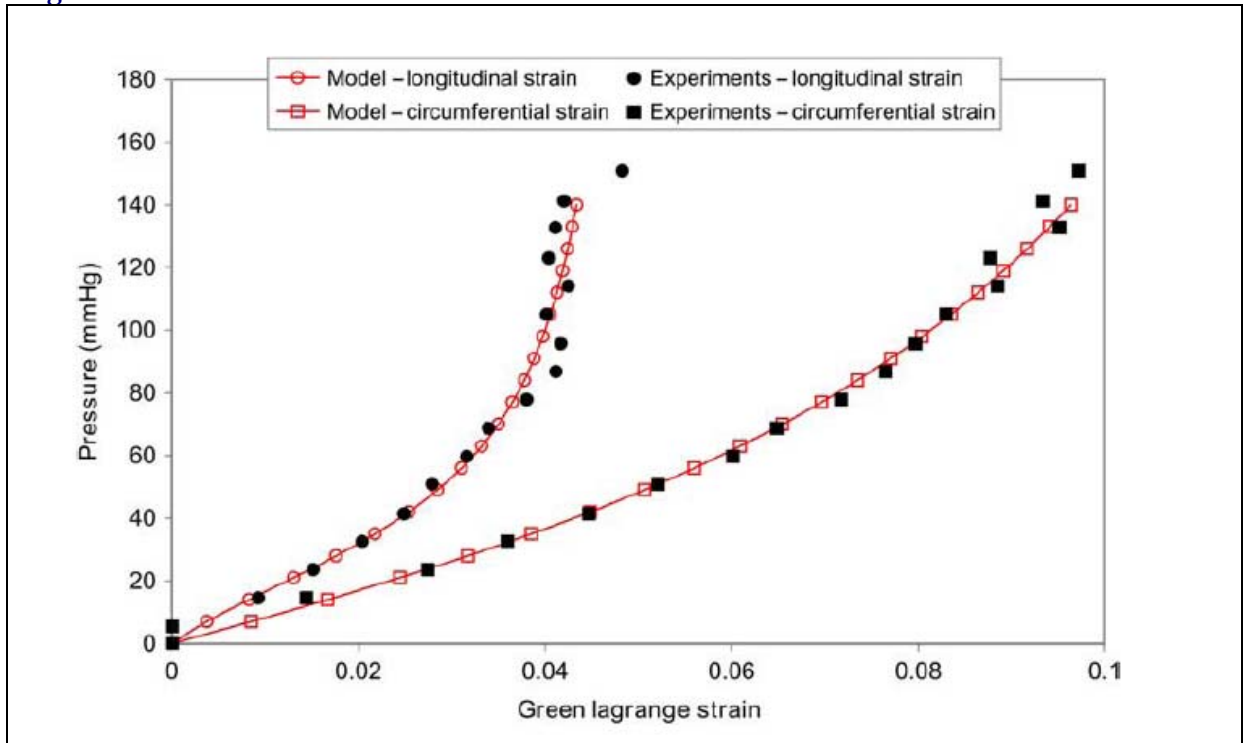

Figure 7:

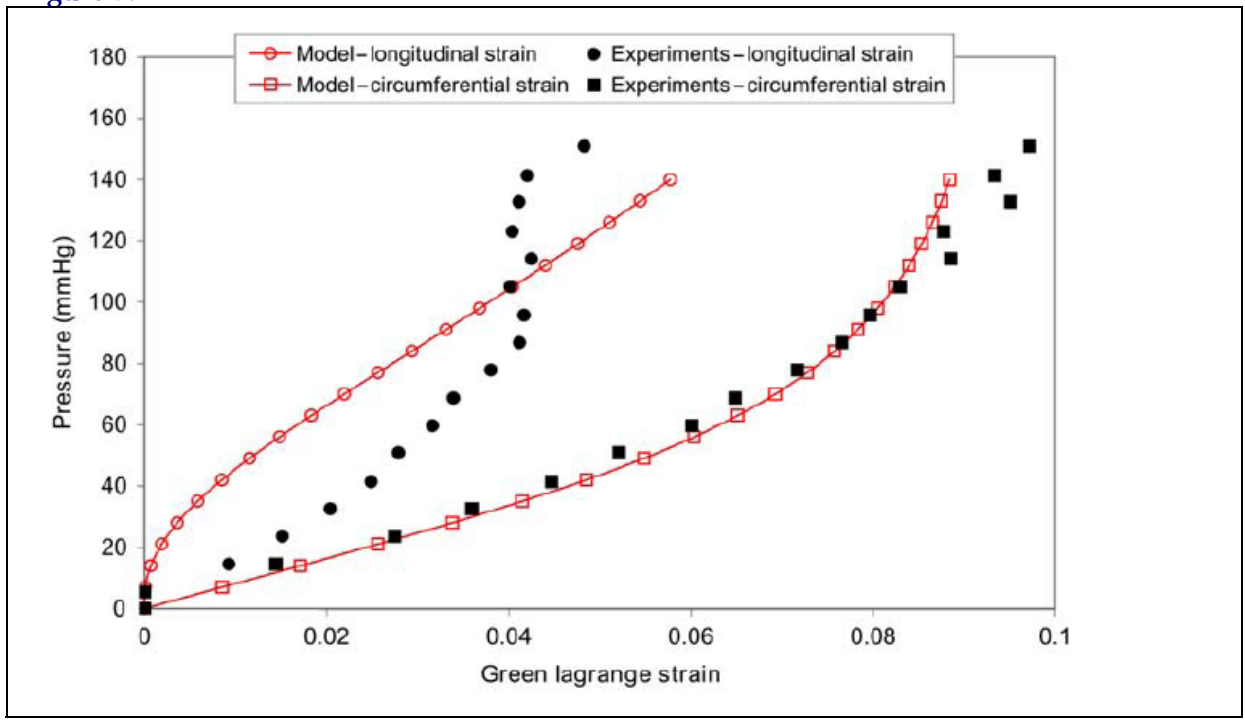